\documentclass[aps,prl,twocolumn]{revtex4-2}
\usepackage{amsfonts}
\usepackage{amssymb}
\usepackage{amsmath}
\usepackage{bbm}
\usepackage{bm}% bold math
\usepackage{relsize}
\usepackage{graphicx}

\usepackage{physics}
\usepackage{textcomp}

\DeclareMathAlphabet\mathbfcal{OMS}{cmsy}{b}{n}

\newcommand{\be}{\begin{equation}}
\newcommand{\ee}{\end{equation}}
\newcommand{\bea}{\begin{eqnarray}}
\newcommand{\eea}{\end{eqnarray}}
\newcommand{\Eq}[1]{Eq.\,(\ref{#1})}% \Eq{abc}

\newcommand{\Fig}[1]{Fig.\,\ref{#1}}% \Fig{fig:abc}

\newcommand{\Sec}[1]{Sec.\,\ref{#1}}% \Sec{sec:abc} sic!byc konsekewntnym \label{sec:xx} \Sec{sec:xx}
\newcommand\wP{\hat{\mathbb P}}

\newcommand\wD{\hat{\mathbb D}}

\newcommand\wF{\vec{\mathbb F}}

\newcommand\one{\hat{\mathbf{1}}}
\newcommand\zero{\hat{\mathbf{0}}}

\newcommand{\heps}{\hat{{\pmb{\varepsilon}}}}

\newcommand{\hxi}{\hat{{\pmb{\xi}}}}
\newcommand{\hzeta}{\hat{{\pmb{\zeta}}}}

\newcommand{\hmu}{\hat{{\pmb{\mu}}}}

\renewcommand{\r}{\textbf{r}}

\newcommand{\E}{\textbf{E}}

\renewcommand{\H}{\textbf{H}}

\renewcommand{\S}{\textbf{S}}

\usepackage{xr} % cross-references to external documents
\begin{document}

\title{
First-order perturbation theory of eigenmodes for systems with interfaces
}
\author{Z. Sztranyovszky}
\author{W. Langbein}
\author{E. A. Muljarov}
\affiliation{School of Physics and Astronomy, Cardiff University, Cardiff CF24 3AA, United Kingdom}
\date{\today}

\begin{abstract}
We present an exact first-order perturbation theory for eigenmodes in systems with interfaces causing material discontinuities. We show that when interfaces deform, higher-order terms of the perturbation series can contribute to the eigenmode frequencies in first order in the deformation depth.
In such cases, the first-order approximation is different from the usual diagonal approximation and its single-mode result. Extracting additional first-order corrections from all higher-order terms enables us to recover the diagonal formalism in a modified form. A general formula for the single-mode first-order correction to electromagnetic eigenmodes in systems with interfaces is derived, capable of treating dispersive, magnetic, and chiral materials of arbitrary shape.

\end{abstract}

\maketitle

Eigenmodes, which are solutions to a  differential equation of Sturm-Liouville type with a set of boundary conditions, are used to describe physical phenomena across physics, including gravitational astronomy~\cite{FerrariGRG08}, acoustics~\cite{KochAIAA05}, seismology~\cite{DahlenBook21}, quantum mechanics (QM)~\cite{ZeldovichSPJETP61}, and electromagnetism (EM)~\cite{AgranovichBook99}. The eigenmodes of open systems are also referred to as resonant states (RSs)~\cite{MuljarovEPL10}, or quasi-normal modes~\cite{YanPRL20}. They determine the optical properties of a resonator, such as its scattering cross-section or Purcell enhancement~\cite{BothSST21}. For simple electromagnetic systems, such as a slab or a sphere, the RSs can be found analytically~\cite{WeinsteinBook69}.
For more complicated shapes they can be found numerically~\cite{YanPRL20,GeOSA14} or via perturbative approaches~\cite{YanPRL20,BothSST21}. The resonant-state expansion (RSE) is a method that treats perturbations in all perturbation orders by transforming the problem of solving Maxwell's equation into a matrix eigenvalue problem~\cite{MuljarovEPL10}. Its accuracy is controlled by the selection of eigenmodes in the basis.

For small changes of the system, it is sufficient to take only a few suited RSs in the basis, or even a single one in a non-degenerate case. The latter corresponds to the diagonal approximation in terms of the matrix equation, and in certain cases this can also be equivalent to the first-order approximation, though not necessarily, as we will show in this paper. Following the terminology of Ref.\,\cite{MorseBook53}, we distinguish two different kinds of perturbations: \textit{volume} perturbation (VP) and \textit{boundary} perturbation (BP). A VP is a small change of the medium properties over a finite volume, for example, in QM a small change in the potential over the width of a quantum well, or in acoustics a small change of the density of the medium. A BP instead moves the spatial position of a medium interface with a discontinuity in medium properties, such as changing the width of a quantum well in QM. In EM, VPs could be a small change of a resonator's permittivity $\Delta \varepsilon$, for which first- and second-order results are well know~\cite{LeungJOSAB96,DoostPRA14,YangACSNL15}, or a change of the medium surrounding the resonator~\cite{BothOL19,AlmousaARXIV22}. The VP examples in EM correspond to the diagonal approximation in the RSE matrix equation, and they include an overlap integral of the eigenmode field with the perturbation (e.g., $\Delta \varepsilon$), in complete analogy with conventional QM~\cite{LandauBook81}. For BPs, this approach is not suited, because the local change of the medium property is not small. Instead, the deformation depth $h(\vb r)$, which is the shift of the surface at position $\vb r$, plays the role of a small perturbation parameter, and an interesting consequence arises from the boundary conditions. For an open system, these are \textit{outgoing waves}~\cite{Peierls76} which cannot be expressed as a combination of Neumann and Dirichlet boundary conditions, which makes the approach of Ref.\,\cite{MorseBook53} inapplicable.
The underlying cause for the different treatment required in EM is the discontinuity of the normal component of the electric field at a material boundary~\cite{HillPRB81}.
We note that similar effects can also occur in condensed matter physics when the effective mass in Schr\"odinger's equation is discontinuous, or in acoustics at the boundary between two media with different densities. In EM, the first-order correction to the RS frequency for a BP was treated for closed isotropic dielectric systems by using the electric displacement field normal to the surface~\cite{JohnsonPRE02}. In case of isotropic open dispersive systems, it was also recognized that the VP diagonal matrix element does not give the correct first-order results, and an alternative treatment was found based on distinguishing the electric fields inside and outside the resonator, and Taylor expanding them~\cite{YanPRL20}. The two approaches~\cite{JohnsonPRE02,YanPRL20} are equivalent apart from the frequency-dependent permittivity and the different field normalization for open systems~\cite{YanPRL20}. It is also known that generally zero-frequency~\cite{LobanovPRA19} or zero-permittivity~\cite{ChenJCP20} longitudinal modes need to be included in the basis alongside the RSs, however the contribution of these additional modes to the first-order results have not been considered in detail. Notably, the above BP methods~\cite{JohnsonPRE02,YanPRL20} do not treat VPs, and the mentioned VP methods do not treat BPs.

The purpose of this Letter is twofold. Firstly, we show that when treating a BP by applying the standard perturbation theory valid for VP, all orders of the perturbation series can contribute linearly in $h$. Using EM for illustration, we demonstrate this surprising finding in terms of static (zero-frequency) modes of a non-dispersive open optical system.

Secondly, we derive a unified treatment of small BP and VP, describing correctly the first-order RS wave number change,  linear in $h$ and $\Delta \varepsilon$, respectively. This treatment is generalized to include frequency dispersion, arbitrary media (including magnetic and chiral), and arbitrary shape. Illustrations for both spherical and non-spherical dispersive systems are provided.

For clarity of presentation, we start by considering an unperturbed dielectric system described by a non-dispersive permittivity tensor $\heps(\r)$ with known RSs having wave numbers $k_n$ and electric fields $\E_n(\r)$. This system is perturbed by a change of the permittivity $\Delta\heps(\r)$. In the RSE approach~\cite{MuljarovEPL10,DoostPRA14}, the electric field $\E(\r)$ of a perturbed RS is expanded as
\be
\E(\r)=\sum_\nu c_\nu\E_\nu(\r)\,,
\label{expansion}
\ee
leading to a matrix eigenvalue problem
\begin{equation}
\label{eq:full_rse}
(k - k_\nu) c_\nu = - k\sum_{\nu'} V_{\nu \nu'} c_{\nu'}\,,
\end{equation}
which determines the exact values of the perturbed RS wave numbers $k$ and the expansion coefficients $c_\nu$ in the limit of all unperturbed modes included in the summation.
Here, index $\nu$ labels both the RSs ($\nu=n$, with $k_n \neq 0$) and static modes ($\nu=\lambda$, with all $k_\lambda = 0$~\cite{LobanovPRA19}), and the matrix elements of the perturbation have the form
\begin{equation}
\label{eq:Vnn}
V_{\nu \nu'} = \int \E_{\nu} (\r) \cdot \Delta\heps(\r) \E_{\nu'} (\r) d \r\,,
\end{equation}
where all fields $\E_{\nu}$ are properly normalized~\cite{MuljarovEPL10,MuljarovOL18}.
From the exact RSE equation~(\ref{eq:full_rse}) one can extract, in the spirit of a standard perturbation theory~\cite{LandauBook81}, corrections to the eigenvalue $k_n$ in all orders, in a form of an infinite series~\cite{DoostPRA14}
\begin{equation}
\label{eq:pert_series}
k = k_n - k_n V_{nn} + k_n V_{nn}^2 +  k_n^2 \sum_{\nu \neq n} \frac{V_{n\nu} V_{\nu n}}{k_n - k_{\nu}}+\dots\,,
\end{equation}
which suggests that
\begin{equation}
\label{eq:first0}
k^{(1)}=- k_n V_{nn}
\end{equation}
is the first-order correction to the wave number $k_n$. In fact, each matrix element \Eq{eq:Vnn} is linear both in the permittivity perturbation $\Delta\heps$ and in the deformation depth $h$ in lowest order.

\begin{figure}[t]
	\centering
	\includegraphics[width=1\linewidth]{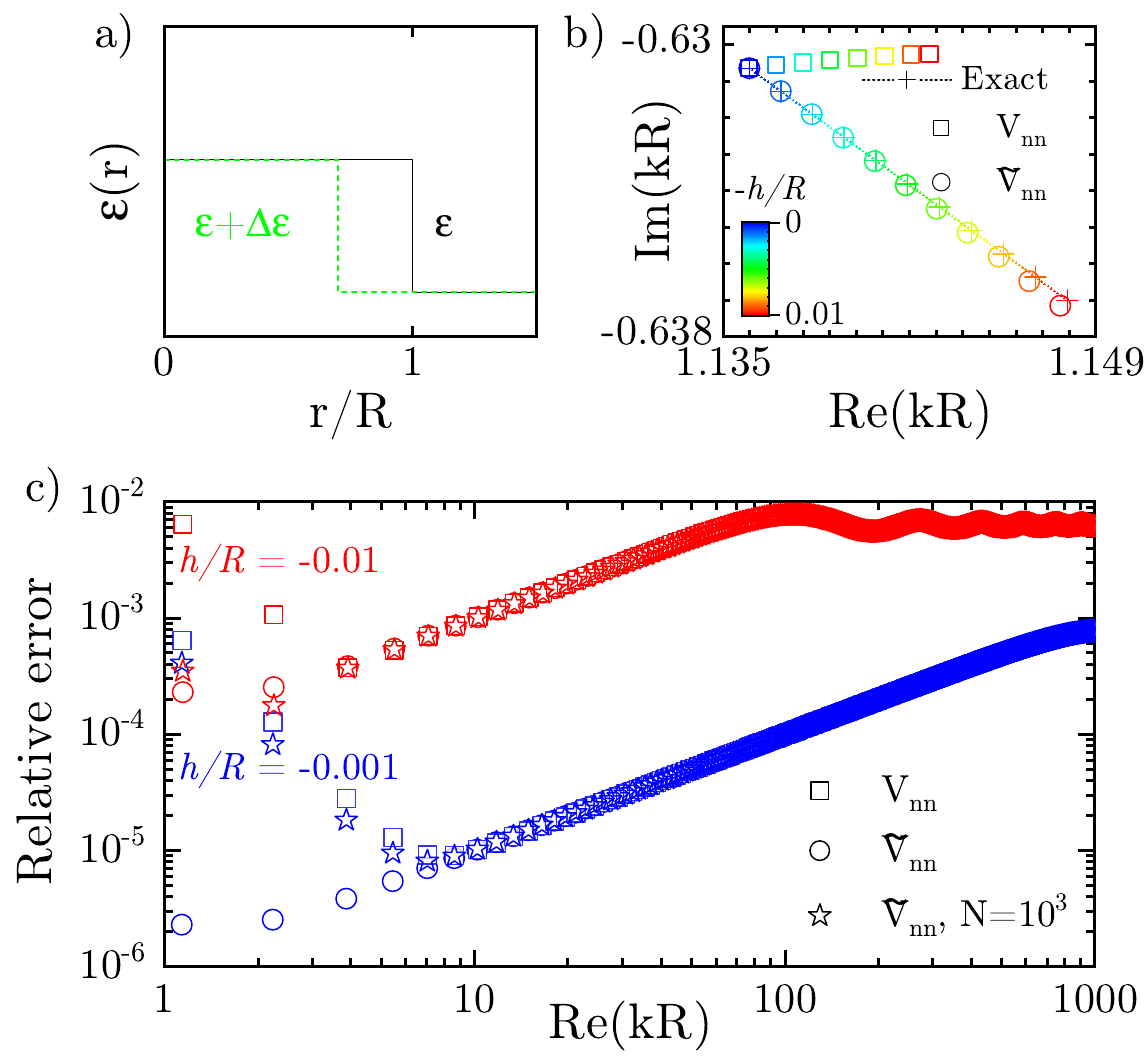}
	\caption{ (a) Schematic illustration of a BP as a permittivity perturbation of the sphere, with its radius $R$ changing by $h$. (b) Effect of the BP on the wave number $k$ of first RS with orbital number $l=1$ in a sphere of permittivity $\varepsilon=4$ surrounded by vacuum. (c) Relative error of RS wave numbers calculated without static modes (squares), with all static modes (circles), and with 1000 static modes included (stars), for two different BPs as given. 
}
\label{f:size}
\end{figure}

The above first-order correction $k^{(1)}$ is illustrated in \Fig{f:size} for TM modes of a dielectric sphere of radius $R$ in vacuum, with angular momentum $l=1$, for a BP changing the radius of the sphere by $h$. Clearly, for the fundamental mode, \Eq{eq:first0} (squares) does not describe correctly the first-order changes of the RS wave number $k$, as a deviation linear in $h$ is observed implying that first-order contributions to $k$ are missing.

The origin of this mismatch lies in the role of static modes, which surprisingly can contribute linearly in $h$ via the second-order sum,  $k_n \sum_\lambda V_{n\lambda} V_{\lambda n}$, and also via all higher-order terms of the perturbation series \Eq{eq:pert_series}. To take their cumulative effect into account, let us write the RSE equation (\ref{eq:full_rse}) in terms of the RSs only, by using the $k_\lambda=0$ degeneracy of static modes~\cite{MuljarovPRA20}:
\begin{equation}
\label{eq:rse}
(k - k_n) c_n = - k\sum_{n'} \tilde V_{nn'} c_{n'}\,,
\end{equation}
where
\begin{equation}
\label{eq:static_matrix_element}
\tilde V_{nn'}=V_{nn'} - \sum_{\lambda \lambda'} V_{n\lambda} W_{\lambda \lambda'} V_{\lambda' n}
\end{equation}
and the matrix $W_{\lambda \lambda'}$ is the inverse of $ \delta_{\lambda \lambda'} + V_{\lambda \lambda'}$ with $\delta_{\lambda \lambda'}$ being the Kronecker delta. The full linear correction to the RS wave number is then given by
\begin{equation}
\tilde k^{(1)}=- k_n \tilde V_{nn}.
\end{equation}

To evaluate the sum in \Eq{eq:static_matrix_element}, we use a Neumann series expansion
$W = (I + V)^{-1} = I - V + V^2 - V^3 + \dots,$
where $W$ ($V$) is a matrix with elements $W_{\lambda \lambda'}$ ($V_{\lambda \lambda'}$) and $I$ is the identity matrix. Substituting it into \Eq{eq:static_matrix_element} results in an infinite series
\begin{equation}
\label{eq:tildeVnn}
\tilde V_{nn'}=V_{nn'} - \sum_{\lambda} V_{n\lambda} V_{\lambda n'}  + \sum_{\lambda \lambda'} V_{n\lambda} V_{\lambda \lambda'} V_{\lambda' n'} -\dots,
\end{equation}
which we evaluate below to first order in $h$, using the static pole residue of the dyadic Green's function (here, its electric part only, which is sufficient for permittivity perturbation). For a spherically symmetric dielectric systems with permittivity $\varepsilon(r)$, the residue can be written explicitly  as (see the SI, \Sec{S-ss:full_rse})
\begin{equation}
\label{eq:residue}
\sum_\lambda \E_\lambda(\r)\otimes \E_\lambda(\r')= \frac{\hat{\r}\otimes \hat{\r}}{\varepsilon(r)} \delta(\r-\r')+\hat{\bf R}(\r,\r')\,,
\end{equation}
where the tensor $\hat{\bf R}(\r,\r')$ is a regular part of the residue, $\hat{\r}$ is the unit vector in the radial direction and $\otimes$ denotes the dyadic product. Using \Eq{eq:residue} for each sum over static modes in \Eq{eq:tildeVnn}, one can see that the $\delta$ function in \Eq{eq:residue} eliminates one volume integration, reducing each term in \Eq{eq:tildeVnn} to a single volume integral, proportional to $h$. Furthermore, the contribution of the regular part can be neglected in linear order in $h$, as it comes with an additional volume integral, and hence is of higher order. Summing over all orders, we arrive, after some algebra, at
\bea
\label{eq:snn_approx}
\frac{\tilde k^{(1)}}{k_n}&=&-\int \E_n  \cdot \left[\one +\frac{\Delta\heps(\r)}{\varepsilon(r)} \hat{\r}\otimes \hat{\r} \right]^{-1} \Delta\heps(\r) \E_n   d\r
\\
&=& -\int \left[\E_n^\parallel \cdot \Delta \varepsilon(\r) \E_n^\parallel + \E_n^\perp \cdot  \frac{\varepsilon(r) \Delta \varepsilon(\r)}{\varepsilon(r) + \Delta \varepsilon (\r)} \E_n^\perp \right] d\r\,,
\nonumber
\eea
where $\one$ is the identity tensor, and the perturbation $\Delta \varepsilon (\r)$  in the second line is assumed isotropic but not necessarily spherically symmetric.
The superscript $\parallel$ ($\perp$) labels the vector component parallel (normal) to the interface of the basis system. More details of the derivation of the above equations and their extension to magnetic and chiral materials are provided in the SI, \Sec{S-s:first_order}.

Figure~\ref{f:size} demonstrates that \Eq{eq:snn_approx} (circles) correctly describes the effect of the size perturbation in first-order.  In fact, comparing $h/R=-0.01$ and $-0.001$, one can see that the residual error scales quadratically, i.e. is of second order in $h$, as expected. We also show in \Fig{f:size} the error for the RS wave numbers calculated with an explicit use of $N=10^3$ static modes~\cite{SztranyovszkyPRA22} in $\tilde V_{nn}$ via \Eq{eq:static_matrix_element} (stars). This demonstrates that as $h$ gets smaller, even a large number of static modes is not sufficient to represent the $\delta$ function in \Eq{eq:residue} well, resulting in errors for the eigenmodes close to zero wave number that are similar with and without the static modes. For higher $l$, the static modes can still contribute to the RSs in first-order, thought the effect is less pronounced due to the higher frequencies of the RSs (see the SI, \Sec{S-s:size}). Note that, while the total number of static modes is countable infinite, the freedom of choosing such a set, granted by their wave number degeneracy, allows one to concentrate the effect of the boundary shift in a single-mode contribution~\cite{ChenJCP21}, which is sufficient to describe the full first-order correction.

As it is clear from \Fig{f:size} and the above derivation, the first-order term of the standard perturbation series \Eq{eq:pert_series} does not contain all first-order effects of the BP. Instead, additional first-order terms can be found in all orders of the perturbation theory. We emphasize that this occurs in any area of physics describing wave phenomena, and we give in the SI, \Sec{S-s:otherfields}, examples from condensed matter and acoustics illustrating this effect.

\begin{figure}[t]
	\centering
	\includegraphics[width=1\linewidth]{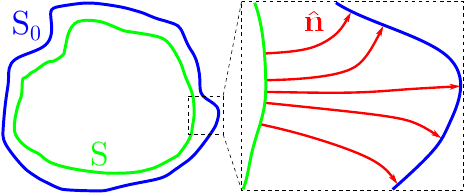}
	\caption{Sketch of the unperturbed and perturbed systems occupying the volumes $V_0$ and $V$ and having surfaces $S_{0}$ and $S$, respectively.
	The vector field $\hat{\bf n}(\r)$ is normal to both $S_{0}$ and $S$.
}
\label{f:sketch}
\end{figure}

The first-order result \Eq{eq:snn_approx} was for clarity obtained for a spherically symmetric non-dispersive system.
We now generalize \Eq{eq:snn_approx} to optical systems with (i) any geometry, (ii) magnetic and chiral materials, and (iii) arbitrary frequency dispersion. To do this, we write Maxwell's equations for the unperturbed system in the compact form~\cite{MuljarovOL18}
\be
\left[k_n\wP_0(k_n,\r)-\wD(\r)\right]\wF_n(\r)=0\,,
\label{ME}
\ee
where
\be
\wP_0=
\begin{pmatrix}
\heps&-i\hxi\\
i\hzeta&\hmu
\end{pmatrix},\ \
\wD=
\begin{pmatrix}
\zero&\nabla\times\\
\nabla\times&\zero
\end{pmatrix},\ \
\wF_n=
\begin{pmatrix}
\E_n\\
i\H_n,
\end{pmatrix}
\label{PD}
\ee
and $\zero$ is the $3\times3$ zero matrix.  $\wP_0(k,\r)$ is a $6\times6$ tensor describing the system which consists of frequency-dispersive tensors of permittivity $\heps(k,\r)$, permeability $\hmu(k,\r)$, and bi-anisotropy $\hxi(k,\r)$ and $\hzeta(k,\r)$. $\wF_n(\r)$ is a $6\times1$ vector comprising $\E_n(\r)$ and $\H_n(\r)$, the electric and magnetic fields of the RS with the wave number $k_n$. Applying a perturbation $\Delta\wP(k,\r)$ of the generalized permittivity, the electromagnetic field and the wave number of this RS change, respectively, to $\wF(\r)$ and $k$, which in turn satisfy perturbed Maxwell's equations
\be
\left[k\wP(k,\r)-\wD(\r)\right]\wF(\r)=0
\label{ME1}
\ee
with $\wP(k,\r)=\wP_0(k,\r)+\Delta\wP(k,\r)$ of the perturbed system. For clarity of presentation, we assume below isotropic and reciprocal materials; anisotropy is considered in the SI, \Sec{S-s:otherfields}, and a further generalization to non-reciprocal materials is possible~\cite{SauvanOE22}. Multiplying \Eq{ME} with $\wF$ and \Eq{ME1} with $\wF_n$,
integrating both equations over the unperturbed system volume $V_0$ (which contains the perturbation $\Delta\wP$, see \Fig{f:sketch}), taking the difference between the results, and applying the divergence theorem to the terms with $\wD$-operators~\cite{MuljarovOL18}, we obtain
\bea
&&\int_{V_0} \wF_n(\r)\cdot \left[k_n\wP_0(k_n,\r)-k\wP(k,\r)\right]\wF(\r) d\r
\nonumber\\
&&=i\oint_{S_{0}}\left[\E_n(\r)\times\H(\r)-\E(\r)\times \H_n(\r)\right]\cdot d\S
\,,
\label{Diff}
\eea
where $S_{0}$ is the boundary of $V_0$.

To extract from \Eq{Diff} the first-order correction $\tilde k^{(1)}$ to the wave number, we introduce a real vector field $\hat{\bf n}(\r)$ which is normal to both surfaces $S_{0}$ and $S$ (of the unperturbed and perturbed systems) and is normalized at each point as $|\hat{\bf n}(\r)|=1$ (if there are other surfaces with material discontinuities, $\hat{\bf n}$ should be chosen normal also to them). Then we introduce a perturbed field component
\be
\wF^\perp(\r)=
\begin{pmatrix}
\E^\perp(\r)\\
i\H^\perp(\r)
\end{pmatrix}
=
\begin{pmatrix}
\hat{\bf n}(\r)[\hat{\bf n}(\r)\cdot \E(\r)]\\
\hat{\bf n}(\r)[\hat{\bf n}(\r)\cdot i\H(\r)]
\end{pmatrix},
\label{Perp}
\ee
which is normal to both $S_{0}$ and $S$. The tangential component is then given by $\wF^\parallel=\wF-\wF^\perp$. Now, according to Maxwell's boundary conditions, fields
$\wF^\parallel(\r)$ and $\wP(k,\r)\wF^\perp(\r)$ are continuous everywhere. Similarly, the unperturbed fields $\wF_n^\parallel(\r)$ and $\wP_0(k_n,\r)\wF_n^\perp(\r)$, introduced in the same manner, are also continuous. Then, approximating $\wF^\parallel(\r)\approx\wF_n^\parallel(\r)$ and $\wP(k,\r)\wF^\perp(\r)\approx\wP_0(k_n,\r)\wF_n^\perp(\r)$, which is sufficient for determining the wave number $k$ to first order, we use in \Eq{Diff}
\be
\wF(\r)= \wF_n^\parallel(\r)+\wP^{-1}(k,\r)\wP_0(k_n,\r)\wF_n^\perp(\r)\,,
\label{WF}
\ee
where $\wP^{-1}$ is the inverse of $\wP$.
Finally, applying a Taylor expansion $k\wP(k,\r)=k_n\wP(k_n,\r)+ [k\wP(k,\r)]'(k-k_n)+\dots$ and $\wF(\r)=\wF_n(\r)+\wF_n'(\r)(k-k_n)+\dots$ for the field outside the systems and keeping only terms linear in $k-k_n$,  we arrive, after some algebra (see the SI, \Sec{S-Sec:Full}), at
\be
\label{eq:first}
\frac{\tilde k^{(1)}}{k_n}=\frac{-\int_{V_0}  \Bigl[\wF^\parallel_n\cdot \Delta\wP\wF^\parallel_n +\wF^\perp_n\cdot \wP_0 \wP^{-1}\Delta\wP\wF^\perp_n\Bigr]d\r}{
\int_{V_0} \!\wF_n\!\cdot \!\bigl[k\wP_0\bigr]'\wF_n d\r\!+\!i\oint_{S_0}\!(\E_n\!\times\!\H_n'\!-\!\E_n'\!\times\!\H_n)\!\cdot\! d\S
},
\ee
where the prime indicates the derivative with respect to $k$ and all quantities are taken at $k=k_n$. Equation (\ref{eq:first}) is a generalization of \Eq{eq:snn_approx}, which is valid also for small perturbations outside the basis system, including deformation outwards.
Note that we have not assumed so far any specific normalization of $\wF_n(\r)$. The analytic normalization introduced in~\cite{MuljarovEPL10,MuljarovOL18} ensures that the denominator in \Eq{eq:first} is equal to 1.

It is important to note that Eqs~(\ref{eq:snn_approx}) and (\ref{eq:first}) contain the exact first-order correction both in terms of the permittivity change ($\Delta\heps$ or $\Delta\wP$) and in the deformation depth $h$. They also include higher order corrections which are not exact. For simplicity we assumed non-degenerate modes in the above derivation. To find the first order correction to degenerate modes, a matrix equation similar to \Eq{eq:rse}, including only degenerate states, will need to be diagonalized.

The above derivation provides a clue for understanding the demonstrated phenomenon that the standard perturbation series \Eq{eq:pert_series} can have contributions to the RS wave number which are linear in $h$ in all perturbation orders. The zeroth-order approximation of the field \Eq{WF}, which is the key point of the derivation, is different from the standard expansion \Eq{expansion} used for a single mode. The failure to extract the correct first order from a series like \Eq{eq:pert_series} technically arises from approximating discontinuous functions with continuous ones. Further illustrations of this fact and a link to the completeness of the basis functions are provided in the SI, \Sec{S-sec:spherical}.

\begin{figure}[t]
	\centering
	\includegraphics[width=1\linewidth]{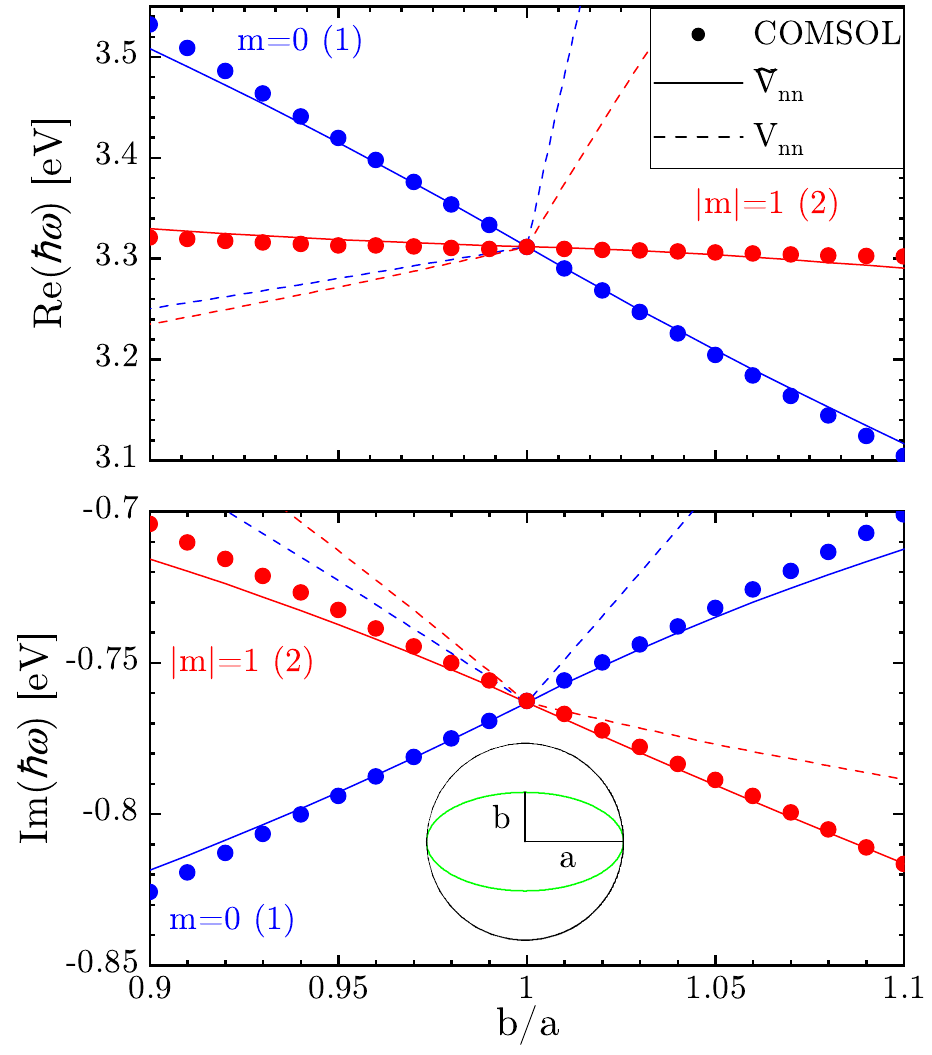}
	\caption{(a) Real and (b) imaginary part of the wave number of the dipolar surface plasmon mode of a silver sphere perturbed to an ellipsoid, which is sketched in the inset of (b). The mode degeneracy is shown in brackets, and $m$ is its magnetic quantum number. The COMSOL data is taken from~\cite{YanPRL20}.
}
\label{f:ellipse}
\end{figure}

We now demonstrate the first-order formula \Eq{eq:first} on a non-spherical system with frequency dispersion.  Figures~\ref{f:ellipse}(a) and (b) show, respectively, the real and imaginary part of $k$ for the dipolar surface plasmon mode of a silver sphere being distorted in to an ellipsoid. The permittivity of silver is given by the Drude model: $\epsilon (\omega)= 1 - \omega_p^2 / (\omega^2 + i\omega \gamma)$ with $\hbar \omega_p = 9$\,eV and $\hbar \gamma = 0.021$\,eV, as used in~\cite{YanPRL20}.
The perturbation theory \Eq{eq:first} (solid lines) agrees in first order of $h$ with numerically calculated values (circles), for both inwards ($a<R$) and outwards ($a>R$) perturbations of the silver sphere. The results using $k^{(1)}$ are also shown for comparison (dashed lines), and are clearly incorrect. This example was chosen identical to the one used in~\cite{YanPRL20}, where the first-order RSE was taken as $k^{(1)}$ given by \Eq{eq:first0}, even though earlier works~\cite{LobanovPRA19,MuljarovPRA20} indicate that static modes could contribute in first order. Thus the statement in Ref.\,\cite{YanPRL20} that the RSE is providing an incorrect first-order result was premature.

In conclusion, we have shown that a first-order perturbation theory of the eigenfrequencies in open systems requires separate considerations for volume perturbations and interface shifts. While volume perturbations lead to first-order diagonal matrix elements capturing the complete first-order effect, moving interfaces which host discontinuities of the underlying medium properties leads to additional first-order contributions arising from higher-order terms. In case of electromagnetism, this is due to the coupling to the countable infinite number of degenerate static modes. The underlying mechanism is clarified by explicitly treating the static pole of the Green’s dyadic, and a first-order perturbation theory expression valid for both medium changes and interface shifts is provided.

\bibliography{references}
\end{document}